\documentclass[intlimits,twoside,a4paper]{article}

\usepackage[cp1251]{inputenc}

\usepackage{xcolor}
\usepackage[eqsecnum]{cmpj3}

\usepackage{bm}

\issue{2018}{21}{1}{13602}
\doinumber{10.5488/CMP.21.13602}

\title[Scaled particle theory for a spherocylinder fluid in a porous medium]%
{Scaled particle theory for a hard spherocylinder fluid in a disordered porous medium:
Carnahan-Starling and Parsons-Lee corrections}
\author[M.F. Holovko, V.I. Shmotolokha]{M.F. Holovko, V.I. Shmotolokha }
\address{
Institute for Condensed Matter Physics of the National Academy of Sciences of Ukraine,\\
1 Svientsitskii St., 79011 Lviv, Ukraine
}

\date{Received February 1, 2018}

\begin{document}

\maketitle

\begin{abstract}
The scaled particle theory (SPT) approximation is
applied for the study of the influence of a porous medium on the
isotropic-nematic transition in a hard spherocylinder fluid. Two new
approaches are developed in order to improve the description in the case
of small lengths of spherocylinders. In one of them, the so-called SPT-CS-PL
approach, the Carnahan-Starling (CS) correction is introduced to
improve the description of thermodynamic properties of the fluid, while
the Parsons-Lee (PL) correction is introduced to improve the
orientational ordering. The second approach, the so-called SPT-PL approach,
is connected with generalization of the PL theory to anisotropic fluids in
disordered porous media. The phase diagram is obtained from the bifurcation
analysis of a nonlinear integral equation for the singlet distribution
function and from the thermodynamic equilibrium conditions. The results
obtained are compared with computer simulation data. Both ways and both
approaches considerably improve the description in the case of spherocylinder
fluids with smaller spherocylinder lengths. We did not find any significant
differences between the results of the two developed approaches. We found
that the bifurcation analysis slightly overestimates and the thermodynamical
analysis underestimates the predictions of the computer simulation data. A
porous medium shifts the phase diagram to smaller densities of the fluid and
does not change the type of the transition.
\pacs 61.20.Gy, 61.43.Gy
\end{abstract}

\section{Introduction}

A hard spherocylinder fluid is one of the simplest popular models widely used
for the description of isotropic-nematic phase transitions in the theory of
liquid crystals \cite{1hol}. During this transition the fluid separates into
two phases with two different densities. The phase with the lower density is
the isotropic one and in this phase the distribution of molecular
orientations is uniform. The other phase with the higher density is the
nematic one and in this phase molecular orientations are strongly ordered.
This phase separation was first explained by Onsager \cite{2hol} nearly
seventy years ago as a result of competition between the orientational
entropy that favours disorder and the entropy effect associated with the
orientational-dependent excluded volume of spherocylinder-like particles that
favours order. Onsager's treatment of the isotropic-nematic transition was given
for a very specific model of a hard spherocylinder fluid in which the length
of spherocylinder $L_1\to\infty $ and the diameter $D_1\to0$ in such a way
that the non-dimensional density of the fluid $c_1=\frac{1}{4}\piup \rho_1
L_1^{2}D_1$ is fixed, where $\rho_1 =\frac{N_1}{V}$, $N_1$ is the number of
spherocylinders, $V$ is the volume of the system. The Onsager theory is based
on the low-density expansion of the free energy functional truncated at the
second virial coefficient level. The result obtained for such a model in this
description is exact \cite{1hol}.

The application of the scaled particle theory (SPT) previously developed for
a hard-sphere fluid \cite{3hol,4hol} provides an efficient approximate way to
incorporate the higher order contributions neglected in the Onsager theory.
As a result, it was possible to generalize the Onsager theory for the
description of a more realistic model of the hard spherocylinder fluid with a
finite value of the length of spherocylinder~$L_1$ and a nonzero value of the
diameter $D_1$ \cite{5hol,6hol,7hol}. We note that for thermodynamic
properties of a hard sphere fluid, the SPT produces the same result as the
Percus-Yevick theory \cite{8hol,9hol}. A basic defect of such a description
is known to appear at higher densities where the theory needs some
improvement, such as a semi-empirical Carnahan-Starling (CS) correction
\cite{10hol,11hol}. Recently the SPT was applied for the description of
isotropic-nematic phase transition in a mixture of hard spheres and hard
spherocylinders. By comparison with the corresponding computer simulation
data \cite{12hol,13hol,14hol} it was shown that the accuracy of the SPT
description reduces with a decreasing length $L_1$ of spherocylinder. Such a
bad accuracy of SPT description can be improved by CS
correction. An alternative to SPT way of improvement of the Onsager theory is
the Parsons-Lee (PL) approach \cite{15hol,16hol,17hol} which is based on the
mapping of the properties of the hard spherocylinder fluid to those of the
hard sphere fluid. The application of the CS theory to the
hard sphere system in the PL theory leads to a correct description of the
isotropic-nematic transition in the hard spherocylinder fluid even at small
lengths $L_1$ of spherocylinders \cite{16hol}.

During the last decade the scaled particle theory was extended to generalize
for the description of a hard sphere fluid in disordered porous media
\cite{18hol,19hol,20hol,21hol,22hol,23hol,24hol}. The obtained results were
generalized for the fluid of hard convex body particles in disordered porous
media \cite{25hol} and was used for the study of the influence of porous
media on the isotropic-nematic transition in a hard spherocylinder fluid in
disordered porous media \cite{26hol,27hol}. It was shown that a porous
medium shifts the isotropic-nematic phase transition to smaller fluid
densities. However, similar to the bulk case, the accuracy of the developed
SPT description reduces with a decreasing spherocylinder length. In this
paper, in order to improve the SPT description of a hard spherocylinder fluid
in disordered porous media we will introduce two types of corrections. The
first one is the CS correction which improves the description
at higher densities of the fluid. The second one corrects the description of
the orientational ordering in a hard spherocylinder fluid at higher
densities. This correction is formulated by comparison of the constants in
the integral equation for the singlet distribution function of hard
spherocylinders in the SPT approach and in the PL theory. In this paper, the
CS and PL corrections constitute the improvement of the
SPT description of a hard spherocylinder fluid in disordered porous media. In
parallel to the SPT approach, we also consider the generalization of the
PL theory for a hard spherocylinder fluid in disordered porous media. It is shown that both approaches provide a correct description of the
isotropic-nematic phase transition in a hard spherocylinder fluid in
disordered porous media including the hard spherocylinder fluids with small
lengths of spherocylinders.

The paper is arranged as follows. In section~\ref{s2} we give a brief review of the
application of the SPT approach for a hard spherocylinder fluid in disordered
porous media. An improvement of the SPT description with the
CS and the PL corrections is presented in section~\ref{s3}. In
section~\ref{s4} generalization of the PL theory for a hard spherocylinder fluid in
disordered porous media is presented. The results and discussion are
presented in section~\ref{s5}. We conclude in section~\ref{s6}.

\section{SPT for hard spherocylinder fluids in disordered porous media \label{s2}}

In this section we present a short review of the SPT for hard spherocylinder
fluids in disordered porous media. The basic idea of the SPT lies in the
insertion of an additional hard spherocylinder with the scaling diameter
$D_{\text s}$ and the scaling length $L_{\text s}$ into a fluid in such a way that
\begin{equation}
D_{\text s}=\lambda_{\text s}D_{1}\,, \quad L_{\text s}=\alpha_{\text s}L_{1}\,,
\label{hol2.1}
\end{equation}
where $D_{1}$ and $L_{1}$ are the diameter and the length of fluid spherocylinder,
respectively. In the presence of porous media, the excess of chemical
potential for the small scaled particle in a spherocylinder fluid confined in
a matrix can be written in the form \cite{26hol}
\begin{align}
\beta\mu_{\text s}^{\text{ex}}&=-\ln
p_{0}(\alpha_{\text s},\lambda_{\text s})-\ln\left\{1-\frac{\eta_{1}}{V_{1}p_{0}(\alpha_{\text s},\lambda_{\text s})}\bigg[\frac{\piup}{6}D_{1}^{3}(1+\lambda_{\text s})^{3}\right.
+\frac{\piup}{4}D_{1}^{2}L_{1}(1+\lambda_{\text s})^{2}(1+\alpha_{\text s})\nonumber
\end{align}
\begin{align}
\label{hol2.2}
&\left.+\frac{\piup}{4}D_{1}L_{1}^{2}(1+\lambda_{\text s})\alpha_{\text s}\int\int
f(\Omega_{1})f({\Omega_{2}})\sin\gamma(\Omega_{1},\Omega_{2})\rd\Omega_{1}\rd\Omega_{2}\bigg]\right\},
\end{align}
where $\beta=\frac{1}{kT}$, $k$ is the Boltzmann constant, $T$ is temperature, $\eta_{1}=\rho_{1}V_{1}$ is the fluid packing
fraction, $\rho_{1}$ is the fluid density, $V_{1}$ is the volume of spherocylinder; $p_{0}(\alpha_{\text s},\lambda_{\text s})$ is the
probability to find a cavity created by a scale particle in the empty matrix
and is defined by the excess of a chemical potential $\mu_{\text s}^{0}$ of the scale
particle in the limit of the infinite dilution of a fluid;
$\Omega=(\vartheta,\varphi)$ is the orientation of particles defined by the
angles $\vartheta$ and $\varphi$; $\rd\Omega=\frac{1}{4\piup}\sin\vartheta \rd
\vartheta \rd \varphi$ is the normalized angle element; $\gamma
(\Omega_{1},\Omega_{2})$ is an angle between orientational vectors of two
molecules; $f(\Omega)$ is the singlet orientational distribution function
normalized in such a way that
\begin{equation}
\int f (\Omega)\rd\Omega=1.
\label{hol2.3}
\end{equation}
For a large scale particle, the excess of chemical potential is given by a
thermodynamic expression that can be presented in the form:
\begin{equation}
\beta\mu_{\text s}^{\text{ex}}=w(\alpha_{\text s},\lambda_{\text s})+{\beta PV_{\text s}} /{ p_{0}(\lambda_{\text s},\alpha_{\text s})},
\label{hol2.4}
\end{equation}
where $P$ is the pressure of the fluid, $V_{\text s}$ is the volume of the scaled
particle, the multiplier $1/{ p_{0}(\lambda_{\text s},\alpha_{\text s})}$ appears due to
an excluded volume confined by matrix particles and can be considered as a
probability to find a cavity created by a scaled particle in the absence of
fluid particles. The probability $p_{0}(\lambda_{\text s},\alpha_{\text s})$ is directly
related to two different types of porosity introduced by us in
\cite{10hol,12hol,14hol,26hol}.

The first one corresponds to the geometrical porosity
\begin{equation}
\phi_0=p_0(\alpha_{\text s}=\lambda_{\text s}=0),
\label{hol2.5}
\end{equation}
characterizing the free volume for a fluid.
The second type of porosity corresponds to the case $\lambda_{\text s}=\alpha_{\text s}=1$ and leads to the thermodynamic porosity
\begin{equation}
\phi=p_0(\alpha_{\text s}=\lambda_{\text s}=1)=\exp(-\beta\mu_{1}^{0}),
\label{hol2.6}
\end{equation}
defined by the excess chemical potential of fluid particles $\mu_{1}^{0}$ in the limit of infinite dilution. It characterizes the adsorption of a
fluid
in the empty matrix. According to the ansatz of the SPT \cite{5hol,6hol,7hol,26hol}
$w(\lambda_{\text s},\alpha_{\text s})$ can be presented in the form:
\begin{equation}
w(\lambda_{\text s},\alpha_{\text s})=w_{{00}}+w_{{10}} \lambda_{\text s}+w_{{01}}\alpha_{\text s}+w_{{11}} \alpha_{\text s}\lambda_{\text s}+\frac {w_{{20}}\lambda_{\text s}^{2}}{2}\,,
\label{hol2.7}
\end{equation}
where the coefficients of this expansion can be found from
 the continuity of the excess chemical potential given in (\ref{hol2.2}) and (\ref{hol2.4}), as well as from the corresponding derivatives
 $\partial\mu_{\text s}^{\text{ex}}/\partial\lambda_{\text s}$, $\partial\mu_{\text s}^{\text{ex}}/\partial\alpha_{\text s}$,
 $\partial^{2}\mu_{\text s}^{\text{ex}}/\partial\alpha_{\text s}\partial\lambda_{\text s}$ and $\partial^{2}\mu_{\text s}^{\text{ex}}/\partial\lambda_{\text s}^{2}$.
As a result, one derives the coefficients as follows:
\begin{equation}
w_{00}=-\ln\left(1-\eta_{1}/\phi_{0}\right),
\label{hol2.8}
\end{equation}
\begin{equation}
w_{10}=\frac{\eta_{1}/\phi_{0}}{1-\eta_{{1}}/\phi_{{0}}}\left(\frac{6\gamma_{1}}{3\gamma_{1}-1}-\frac{p'_{0\lambda}}{\phi_0}\right),
\label{hol2.9}
\end{equation}
\begin{equation}
w_{01}=\frac{\eta_{{1}}/\phi_{0}}{1-\eta_{1}/ \phi_{0}}\left[\frac{3(\gamma_{1}-1)}{3\gamma_{1}-1}
+\frac{3(\gamma_{1}-1)^2}{3\gamma_{1}-1}\tau(f)-\frac{p'_{0\alpha}}{\phi_0}\right],
\label{hol2.10}
\end{equation}
\begin{eqnarray}
w_{11}&=&\frac{\eta_{1}/\phi_{0}}{1-\eta_{1}/\phi_{0}}\Biggl[\frac{6(\gamma_{1}-1)}{3\gamma_{1}-1}+\frac{3(\gamma_{1}-1)^2\tau(f)}{3\gamma_{1}-1}
        -\frac{p''_{0\alpha\lambda}}{\phi_{0}}\nonumber\\
         &+&2\frac{p'_{0\alpha}p'_{0\lambda}}{\phi_{0}^{2}}-\frac{3(\gamma_{1}-1)+3(\gamma_{1}-1)^{2}\tau(f)}{3\gamma_{1}-1}
        \frac{p'_{0\lambda}}{\phi_{0}}
        -\frac{6\gamma_{1}}{3\gamma_{1}-1}\frac{p'_{0\alpha}}{\phi_{0}}\Biggr] \nonumber\\
        &+&\left(\frac{\eta_{1}/\phi_{0}}{1-\eta_{1}/\phi_{0}}\right)^{2}\left(\frac{6\gamma_{1}}{3\gamma_{1}-1}
        -\frac{p'_{0\lambda}}{\phi_0}\right)
        \left[\frac{3(\gamma_{1}-1)}{3\gamma_{1}-1}
         +\frac{3(\gamma_{1}-1)^{2}\tau(f)}{3\gamma_{1}-1}-\frac{p'_{0\alpha}}{\phi_0}\right],
         \label{hol2.11}
        \end{eqnarray}
\vspace{-6mm}
\begin{eqnarray}
\nonumber w_{20}&=&\frac{\eta_{1}/\phi_{0}}{ 1-\eta_{{1}}/\phi_{0}}
\left[\frac{6\left(1+\gamma_{1}\right)}{3\gamma_{1}-1}-
\frac {12\gamma_{1}}{3\gamma_{1}-1}  \frac{p'_{0\lambda}}{ \phi_{0}}+
2\left(\frac{p'_{0\lambda}}{\phi_{0}}\right)^{2}-
\frac{p''_{0\lambda\lambda}}{\phi_{0}}\right]\\
&+&\left(\frac{\eta_{1}/\phi_{0}}{1-\eta_{1}/\phi_{0}} \right)^2
\left(\frac{6\gamma_{1}}{3\gamma_{1}-1}-\frac{p'_{0\lambda}}{\phi_0}\right)^{2},
\label{hol2.12}
\end{eqnarray}
where
\begin{eqnarray}
&\gamma_{1}=1+\frac{L_{1}}{D_{1}}\,,&\\
\label{hol2.13}
%\end{eqnarray}
%%
%\begin{eqnarray}
&\tau(f)=\frac{4}{\piup}\int\int f(\Omega_{1}) f(\Omega_{2}) \sin \gamma(\Omega_{1},\Omega_{2}) \rd\Omega_{1} \rd\Omega_{2}.&
\label{hol2.14}
\end{eqnarray}
Setting $\alpha_{\text s}=\lambda_{\text s}=1$ in the equation~(\ref{hol2.4}) leads to
the expression
\begin{align}
\beta\left(\mu_{1}^{\text{ex}}-\mu_{1}^{0}\right)=-\ln \left(1-\eta_{1}/\phi_{0}\right)+A\big(\tau(f)\big)\frac{\eta_{1}/\phi_{0}}{1-\eta_{1}/\phi_{0}}
+B\big(\tau(f)\big)\frac{(\eta_{1}/\phi_{0})^{2}}{(1-\eta_{1}/\phi_{0})^{2}}\,,
\label{hol2.15(1)}
\end{align}
where the coefficients $A\big(\tau(f)\big)$ and $B\big(\tau(f)\big)$ define the porous medium
structure and the expressions for them are as follows:
\begin{eqnarray}
A\big(\tau(f)\big) &=&6+\frac{6\left(\gamma_{1}-1\right)^2\tau(f)}{3\gamma_{1}-1}-
   \frac{p'_{0\lambda}}{\phi_0}\left[4+\frac{3\left(\gamma_{1}-1\right)^2\tau(f)}{3\gamma_{1}-1}\right] \nonumber\\
&-&\frac{p'_{0\alpha}}{\phi_0}\left(1+\frac{6\gamma_{1}}{3\gamma_{1}-1}\right)-\frac{p''_{0\alpha\lambda}}{\phi_0}-\frac{1}{2}
    \frac{p''_{0\lambda\lambda}}{\phi_0}+2\frac{p'_{0\alpha}p'_{0\lambda}}{\phi_0^{2}}+\left(\frac{p'_{0\lambda}}{\phi_0}\right)^2,
\label{hol2.15}
\end{eqnarray}
 \begin{eqnarray}
B\big(\tau(f)\big) =\left(\frac{6\gamma_{1}}{3\gamma_{1}-1}-\frac{p'_{0\lambda}}{\phi_0}\right)\left[\frac{3\left(2\gamma_{1}-1\right)}{3\gamma_{1}-1}+
\frac{3\left( \gamma_{1}-1\right)^2\tau(f)}{3\gamma_{1}-1}-\frac{p'_{0\alpha}}{\phi_0}
-\frac{1}{2}\frac{p'_{0\lambda}}{\phi_0}\right],
\label{hol2.16}
 \end{eqnarray}
where
 $ p'_{0\lambda } = \frac{\partial {p_0}(\alpha_{\text s},\lambda_{\text s})}{\partial \lambda_{\text s}}$,
 $p'_{0\alpha } =  \frac{\partial p_{0}(\alpha_{\text s} ,\lambda_{\text s})}{\partial \alpha_{\text s}}$,
 $ p''_{0\alpha\lambda } =  \frac{\partial^{2} {p_0}(\alpha_{\text s},\lambda_{\text s})}{\partial\alpha_{\text s}\partial \lambda_{\text s}}$, 
 $p''_{0\lambda\lambda } = \frac{\partial^2{p_0}(\alpha_{\text s},\lambda_{\text s})}{\partial
\lambda_{\text s}^{2}}$ are the corresponding derivatives at $ \alpha=\lambda=0$.
Using the Gibbs-Duhem equation $\left(\frac{\partial P}{\partial\rho_{1}}\right)_{T}=\rho_{1}\left(\frac{\partial
\mu_{1}}{\partial\rho_{1}}\right)_{T}$, which relates the pressure $P$ of a fluid to its total chemical potential
$\mu_{1}=\ln(\eta_{1})+\mu_{1}^{0}+\mu_{1}^{\text{ex}}$
one derives the fluid compressibility in the form
\begin{align}
&\beta\left(\frac{\partial P}{\partial\rho_{1}}\right)_{T}=\frac{1}{\left(1-\eta_{1}/\phi\right)}
+\big[1+A\big(\tau(f)\big)\big]\frac{\eta_{1}/\phi_{0}}
{\left({1-\eta_{1}/\phi}\right)
\left(1-\eta_{1}/\phi_{0}\right)}\nonumber\\
&+\big[A\big(\tau(f)\big)+2B\big(\tau(f)\big)\big]\frac{\left(\eta_{1}/\phi_{0}\right)^{2}}{\left(1-\eta_{1}/\phi\right)\left(1-\eta_{1}/\phi_{0}\right)^{2}}
+2B\big(\tau(f)\big)\frac{\left(\eta_{1}/\phi_{0}\right)^{3}}{\left(1-\eta_{1}/\phi\right)\left(1-\eta_{1}/\phi_{0}\right)^{3}}.\label{hol2.17}
\end{align}

From expression~(\ref{hol2.17}) it is possible to obtain the chemical
expression and the pressure of the fluid in SPT2 approach
\cite{10hol,12hol,26hol}. The expression~(\ref{hol2.17}) at higher densities has two
divergences, which appear at $\eta_{1}=\phi$ and $\eta_{1}=\phi_{0}$,
respectively. Since $\phi<\phi_{0}$, the divergence at $\eta_{1}=\phi$ occurs at
lower densities than the second one and, therefore, should be removed.
Different corrections and improvements of SPT2 approach were proposed in
\cite{10hol,11hol,12hol,14hol,26hol}. The first corrections were considered
in \cite{10hol} where based on SPT2, four different approximations  were proposed. The best one is the SPT2b approximation which was
derived replacing $\phi$ by $\phi_{0}$ everywhere in (\ref{hol2.17}) except
the first term. However, this term has a divergence at $\eta_{1}=\phi$ and due
to this, some other approximations were proposed in
\cite{11hol,12hol,14hol,26hol}. One of them called SPT2b1 can be obtained
from the expression for the chemical potential in SPT2b approach by removing
the divergence at $\eta_{1}=\phi$ through the expansion of the logarithmic
term in the SPT2b expression for the chemical potential as follows:
 \begin{equation}
 -\ln(1-\eta_{1}/\phi)\approx-\ln(1-\eta_{1}/\phi)+\frac{\eta_{1}(\phi_0-\phi)}{\phi_0 \phi(1-\eta_{1}/\phi_{0})}.
 \label{hol2.18}
 \end{equation}

Therefore, one obtains the expressions for the chemical potential and pressure within the SPT2b1 approximation as follows:
\begin{align}
	&\left[\beta\big(\mu_{1}^{\text{ex}}-\mu_{1}^{0}\big)\right]^{\text{SPT2b1}}=\sigma(f)-\ln(1-\eta_{1}/\phi_{0})+\big[1+A\big(\tau(f)\big)\big]\frac{\eta_{1}/\phi_{0}}{1-\eta_{1}/\phi_{0}}
+\frac{\eta_{1}(\phi_{0}-\phi)}{\phi_{0}\phi(1-\eta_{1}/\phi_{0})}
\nonumber\\
&
+\frac12\big[A\big(\tau(f)\big)+2B\big(\tau(f)\big)\big]\frac{(\eta_{1}/\phi_{0})^{2}}{(1-\eta_{1}/\phi_{0})^{2}}+\frac{2}{3}B\big(\tau(f)\big)\frac{(\eta_{1}/\phi_{0})^{3}}
{(1-\eta_{1}/\phi_{0})^{3}}\,,
\label{hol2.19}
\end{align}
\begin{align}
&\left(\frac{\beta P}{\rho_{1}}\right)^{\text{SPT2b1}}=\frac{1}{1-\eta_{1}/\phi_{0}}\frac{\phi_{0}}{\phi}+\left(\frac{\phi_{0}}{\phi}-1\right)
\frac{\phi_{0}}{\eta_{1}}\ln(1-\eta_{1}/\phi_{0})\nonumber\\
&+\frac{A\big(\tau(f)\big)}{2}\frac{\eta_{1}/\phi_{0}}{(1-\eta_{1}/\phi_{0})^{2}}+\frac{2B\big(\tau(f)\big)}{3}\frac{(\eta_{1}/\phi_{0})^{2}}{(1-\eta_{1}/\phi_{0})^{3}}\,,
\label{hol2.20}
\end{align}
where
\begin{eqnarray}
\sigma(f)=\int f(\Omega)\ln f(\Omega)\rd\Omega
\label{hol2.21}
\end{eqnarray}
is the entropic term.

Some other approximations which include the third type of porosity $\phi^{*}$
defined by the maximum value of packing fraction of a fluid in a porous media
are analyzed in \cite{11hol,12hol,14hol}. However, in this paper we 
restrict our consideration to the SPT2b1 approximation which is quite
accurate at small, intermediate and higher fluid densities.

From the thermodynamic relationship
\begin{eqnarray}
\frac{\beta F}{V}=\beta \mu_1\rho_1-\beta P,
\label{hol2.22}
\end{eqnarray}
one can obtain an expression for the free energy. Within the SPT2b1
approximation, the free energy of a confined fluid is as follows:
\begin{align}
& \left(\frac{\beta F}{N}\right)^{\text{SPT2b1}} =  \sigma(f)+\ln\frac{\eta_{1}}{\phi}-1
-\ln(1-\eta_{1}/\phi_{0})+\left(1-\frac{\phi_{0}}{\phi} \right)\bigg[1+\frac{\phi_{0}}{\eta_1}\ln(1-\eta_1/\phi_{0})\bigg] \nonumber\\
&+ \frac{A\big(\tau(f)\big)}{2} \frac{\eta_1/\phi_0}{1-\eta_1/\phi_0}+ \frac{B\big(\tau(f)\big)}{3}
\left(\frac{\eta_1/\phi_0}{1-\eta_1/\phi_0}\right)^2.
\label{hol2.23}
\end{align}
The singlet orientational distribution function $f(\Omega)$ can be obtained
from the minimization of the free energy with respect to variations of this
distribution. This procedure leads to the nonlinear integral equation
\begin{equation}
\ln f(\Omega_1)+\lambda+\frac{8}{\piup}C\int f(\Omega')\sin\gamma(\Omega_1\Omega')\rd\Omega'=0,
\label{hol2.24}
\end{equation}
where
\begin{align}
C^{\text{SPT2b1}}=\frac{\eta_{1}/\phi_{0}}{1-{\eta_{1}/\phi_{0}}}\left[\frac{3(\gamma_{1}-1)^{2}}{3\gamma_{1}-1}\left(1-\frac{p'_{0\lambda}}{2\phi_{0}}\right)
+\frac{{\eta_{1}/\phi_{0}}}{(1-{\eta_{1}/\phi_{0}})}\frac{(\gamma_{1}-1)^{2}}{3\gamma_{1}-1}\left(\frac{6\gamma_{1}}{3\gamma_{1}-1}
-\frac{p'_{0\lambda}}{\phi_{0}}\right)\right].
\label{hol2.25}
\end{align}

\section{Carnahan-Starling and Parsons-Lee corrections \label{s3}}
As it was already noted at the beginning of this paper, the SPT approach is
not accurate enough for higher fluid densities as the length of
spherocylinders decreases and the CS correction \cite{11hol}
should be taken into account. The CS correction is generalized
for the presence of a porous media. We present the equation of state in the
following form:
\begin{eqnarray}
\left(\frac{\beta P}{\rho_{1}}\right)^{\text{SPT2b1-CS}}=\left(\frac{\beta P}{\rho_{1}}\right)^{\text{SPT2b1}}+\left(\frac{\beta \Delta P}{\rho_{1}}\right)^{\text{CS}},
\label{hol3.1}
\end{eqnarray}
where $({\beta P}/{\rho_{1}})^{\text{SPT2b1}}$ is given by equation~(\ref{hol2.20}), $({\beta \Delta P}/{\rho_{1}})^{\text{CS}}$ is the CS correction
which we present in the form
\begin{eqnarray}
\left(\frac{\beta \Delta P}{\rho_{1}}\right)^{\text{CS}}=-\frac{\left(\eta_{1}/\phi_0\right)^3}{\left(1-\eta_{1}/\phi_0\right)^3}.
\label{hol3.2}
\end{eqnarray}
We present the chemical potentials in a similar form
\begin{eqnarray}
\label{hol3.3}
(\beta\mu_{1})^{\text{SPT2b1-CS}}=(\beta\mu_{1})^{\text{SPT2b1}}+(\beta\Delta\mu_{1})^{\text{CS}},
\end{eqnarray}
where the correction $(\Delta\mu_{1})^{\text{CS}}$ can be obtained from the Gibbs-Duhem
equation
\begin{eqnarray}
\label{hol3.4}
(\beta\Delta\mu_{1})^{\text{CS}}=\beta\int_{0}^{\eta_{1}}\frac{\rd\eta_{1}}{\eta_{1}} \left(\frac{\partial \Delta P}{\partial \rho_{1}}\right)^{\text{CS}}.
\end{eqnarray}
As a result,
\begin{eqnarray}
\label{hol3.5}
(\beta\Delta\mu_{1})^{\text{CS}}=\ln(1-\eta_{1}/\phi_0)+\frac{\eta_{1}/\phi_0}{1-\eta_{1}/\phi_0}-\frac{1}{2}\frac{(\eta_{1}/\phi_0)^{2}}{(1-\eta_{1}/\phi_0)^{2}}-\frac{(\eta_{1}/\phi_0)^{3}}{(1-\eta_{1}/\phi_0)^{3}}.
\end{eqnarray}
The free energy can also be presented in the form

\begin{eqnarray}
\label{hol3.6}
\left(\frac{\beta F}{N_1}\right)^{\text{SPT2b1-CS}}=\left(\frac{\beta F}{N_1}\right)^{\text{SPT2b1}}+\left(\frac{\beta F}{N_1}\right)^{\text{CS}},
\end{eqnarray}
where the first term $({\beta F}/{N_{1}})^{\text{SPT2b1}}$ is given by
equation~(\ref{hol2.23}) and the second term can be found from thermodynamic
relation~(\ref{hol2.22})
\begin{eqnarray}
\label{hol3.7}
\left(\frac{\beta \Delta
F}{N_{1}}\right)^{\text{CS}}=\ln(1-\eta_{1}/\phi_0)+\frac{\eta_{1}/\phi_0}{1-\eta_{1}/\phi_0}-\frac{1}{2}\frac{\left(\eta_{1}/\phi_0\right)^2}{\left(1-\eta_{1}/\phi_0\right)^2}.
\end{eqnarray}

However, the considered CS correction improves only
thermodynamic properties and does not modify the description of orientational
ordering which is described by the integral equation~(\ref{hol2.24}) for the
singlet distribution function $f(\Omega)$. In order to improve the
description of orientational ordering we should modify the parameter $C$
given by the expression~(\ref{hol2.25}). This parameter has two terms. The
first term appears from the coefficient $A\big(\tau(f)\big)$ and the second one
appears from the coefficient $B\big(\tau(f)\big)$ in the expression~(\ref{hol2.23})
for the free energy. By simple comparison of parameter $C$ in the SPT
approach and the PL theory for the bulk case we can see that the first term
which appears from the coefficient $A\big(\tau(f)\big)$ in SPT2b1 theory is the same
as in the PL approach. Although there are some differences in the second term which
appears from the coefficient $B\big(\tau(f)\big)$, it is possible to have practically the same result for the description of isotropic-nematic 
transition from SPT and PL approaches if we introduce some parameter~$\delta$ as a multiplier near the term with $\tau(f)$ in
coefficient $B\big(\tau(f)\big)$. After generalization of this result for the hard spherocylinder
fluid in disordered porous media, we can rewrite the expression for
$B\big(\tau(f)\big)$ in the following form
\begin{eqnarray}
B\big(\tau(f)\big) =\left(\frac{6\gamma_{1}}{3\gamma_{1}-1}-\frac{p'_{0\lambda}}{\phi_0}\right)\left[\frac{3\left(2\gamma_{1}-1\right)}{3\gamma_{1}-1}+
\frac{3\left( \gamma_{1}-1\right)^2\delta\tau(f)}{3\gamma_{1}-1}-\frac{p'_{0\alpha}}{\phi_0}
-\frac{1}{2}\frac{p'_{0\lambda}}{\phi_0}\right].
\label{hol3.8}
\end{eqnarray}
Using the Parsons-Lee approach in the framework of Onsager investigation for
sufficiently long spherocylinders we determined that $\delta=3/8$
\cite{16hol}. As a result, we can present the constant $C$ in the form
\begin{align}
C^{\text{CS-PL}}&=\frac{\eta_{1}/\phi_{0}}{1-{\eta_{1}/\phi_{0}}}\left[\frac{3(\gamma_{1}-1)^{2}}{3\gamma_{1}-1}\left(1-\frac{p'_{0\lambda}}{2\phi_{0}}\right)+\frac{{\eta_{1}/\phi_{0}}}{(1-{\eta_{1}/\phi_{0}})}\delta\frac{(\gamma_{1}-1)^{2}}{3\gamma_{1}-1}\left(\frac{6\gamma_{1}}{3\gamma_{1}-1}
-\frac{p'_{0\lambda}}{\phi_{0}}\right)\right].
\label{hol3.9}
\end{align}

\section{Generalization of the Parsons-Lee theory for the hard spherocylinder fluid in disordered porous media \label{s4}}

In this section we generalize the PL theory for the case of hard
spherocylinder fluid in disordered porous media. In \cite{16hol} in the
framework of the functional scaling concept, a direct generalization of the
CS equation for the free energy of hard sphere fluid for a
nematic fluid was constructed. Following \cite{16hol}, in accordance with
(\ref{hol2.24}) and (\ref{hol3.7}), a generalized expression for the hard
spherocylinder fluid in disordered porous media can be written as
\begin{align}
&\left(\frac{\beta  F}{N_{1}}\right)^{\text{PL}}= \ln\left( \frac{\eta_{1}}{\phi}\right) -1+\sigma(f) +\bigg\{ \left(1-\frac{\phi_0}{\phi}\right)\left[1+\frac{\phi_0}{\eta_1}\ln(1-
\phi_0/\eta_1)\right]\nonumber\\  &+ \left( 1+ \frac{A}{2}\right)   \frac{\eta_1/\phi_0}{1-\eta_1/\phi_0}+\left(  \frac{B}{3}-\frac{1}{2}\right)
\left(\frac{\eta_1/\phi_0}{1-\eta_1/\phi_0}\right)^2\bigg\} \left[1+\frac{3}{4}  \frac{(\gamma_{1}-1)^2}{3\gamma_{1}-1}\tau(f)\right] ,
\label{hol4.1}
\end{align}
where
\begin{eqnarray}
A=6+4\frac{p'_{0\lambda}}{\phi_0}+\left( \frac{p'_{0\lambda}}{\phi_0}\right) ^2-\frac{1}{2}\frac{p''_{0\lambda}}{\phi_0}\,,
\label{hol4.2}
\end{eqnarray}
\begin{eqnarray}
B=\frac{1}{2}\left( 3-\frac{p'_{0\lambda}}{\phi_0}  \right)^2.
\label{hol4.3}
\end{eqnarray}

For the pressure and the chemical potential we will respectively have 
\begin{align}
&\left(\frac{\beta
P}{\rho_{1}}\right)^{\text{PL}}=1+\Bigg\{\frac{\phi_{0}}{\phi}\frac{\eta_{1}/\phi_{0}}{1-\eta_{1}/\phi_{0}}+\bigg(\frac{\phi_{0}}{\phi}-1\bigg)\bigg[ 1+
\frac{\phi_{0}}{\eta_{1}}\ln(1-\eta_{1}/\phi_{0})\bigg] +\frac{A}{2}\frac{\eta_{1}/\phi_{0}}{(1-\eta_{1}/\phi_{0})^{2}}\nonumber\\
&+\frac{2B}{3}\frac{(\eta_{1}/\phi_{0})^{2}}{(1-\eta_{1}/\phi_{0})^{3}}-
\frac{(\eta_{1}/\phi_{0})^3}{(1-\eta_{1}/\phi_{0})^3}\Bigg\}\left[1+\frac{3}{4}  \frac{(\gamma_{1}-1)^2}{3\gamma_{1}-1}\tau(f)\right],
\label{hol4.4}
\end{align}
\begin{align}
&\left(\beta\mu_{1}\right)^{\text{PL}}=\ln\left( \frac{\eta_{1}}{\phi}\right) +  \sigma(f)+\bigg\{\bigg( 1+\frac{\phi_{0}}{\phi}+A\bigg)
\frac{\eta_{1}/\phi_{0}}{1-\eta_{1}/\phi_{0}}+\left[ \frac{1}{2}(A-1)+B\right]  \frac{(\eta_{1}/\phi_{0})^2}{(1-\eta_{1}/\phi_{0})^2}\nonumber\\
&+\left( \frac{2B}{3}-1\right) \frac{(\eta_{1}/\phi_{0})^{3}}{(1-\eta_{1}/\phi_{0})^{3}}\bigg\}\left[1+\frac{3}{4}
\frac{(\gamma_{1}-1)^2}{3\gamma_{1}-1}\tau(f)\right].
\label{hol4.5}
\end{align}
After minimization of the free energy, in the considered approach we obtain
an integral equation for the singlet orientational distribution function in
the form (\ref{hol2.24}), in which, however, the constant $C$ has the following
form
\begin{align}
C^{\text{PL}}&=\frac{6}{\piup}\frac{(\gamma_{1}-1)^2}{3\gamma_{1}-1}
\bigg\{\left(1-\frac{\phi_{0}}{\phi}\right)\left[1+\frac{\phi_{0}}{\eta_{1}}\ln(1-\eta_{1}/\phi_{0})\right]\left( 1+ \frac{A}{2}\right)   \frac{\eta_1/\phi_0}{1-\eta_1/\phi_0}\nonumber\\
&+\left(  \frac{B}{3}-\frac{1}{2}\right)
\left(\frac{\eta_1/\phi_0}{1-\eta_1/\phi_0}\right)^2\bigg\}.
\label{hol4.6}
\end{align}

\section{Results and discussions \label{s5}}

We will illustrate the developed approaches for the hard spherocylinder fluid
in a hard sphere matrix. First, we specify the geometrical and the
probe particle porosities \cite{26hol}. The geometrical porosity $\phi_{0}$
in this case has the form
\begin{align}
\phi_{0}=1-\eta_{0} \,,
\label{hol5.1}
\end{align}
where $\eta_{0}=\rho_{0}V_{0}$, $\rho_{0}=\frac{N_{0}}{V}$, $N_{0}$ is the
number of matrix particles, $V_{0}=\frac{1}{6}\piup D_{0}^3$ is the volume of a
matrix particle, $V$ is the total volume of the system, $D_{0}$ is the
diameter of matrix hard spheres.

Using the SPT, the following expression for the probe particle porosity is
derived \cite{26hol}
 \begin{align}
\phi&=(1-\eta_{0})\exp\left\{-\frac{\eta_{0}}{1-\eta_{0}}\frac{D_{1}}{D_{0}}\left[\frac32(\gamma_{1}+1)+3\gamma_{1}\frac{D_{1}}{D_{0}}\right]-\frac{\eta_{0}^{2}}{(1-\eta_{0})^{2}}\frac92\gamma_{1}\frac{D_{1}^{2}}{D_{0}^{2}}\right.\nonumber\\&\left.-\frac{\eta_{0}^{3}}{(1-\eta_{0})^{3}}(3\gamma_{1}-1)\frac12\frac{D_{1}^{3}}{D_{0}^{3}}\big(1+\eta_{0}+\eta_{0}^{2}\big)\right\}.
\label{hol5.2}
\end{align}
The probability to find a small scaled spherocylinder in an empty matrix is
\begin{equation}
p_{0}(\alpha_{\text s},\lambda_{\text s})=1-\eta_{0}\frac{1}{V_{0}}\frac{\piup}{2}\left[\frac13(D_{0}+\lambda_{\text s} D_{1})^{3}+\frac12\alpha_{\text s} L_{1}
(D_{0}+\lambda_{\text s} D_{1})^{2}\right].
\label{hol5.3}
\end{equation}

Hereupon we can find the derivatives needed for the description of thermodynamic properties of a confined fluid:
\begin{align}
p'_{0\lambda}=-3\frac{D_{1}}{D_{0}}\eta_{0}\,,\qquad p'_{0\alpha}&=-\frac{3}{2}\eta_{0}\frac{L_{1}}{D_{0}}\,,\qquad
p''_{0\alpha\lambda}=-3\eta_{0}\frac{L_{1}}{D_{0}}\frac{D_{1}}{D_{0}}\,, \qquad
p''_{0\lambda\lambda}=-6\eta_{0}\frac{D_{1}^{2}}{D_{0}^{2}}.
\label{hol5.4}
\end{align}

Now we apply the theory presented in the previous section for investigation
of the isotropic-nematic phase transition in a hard spherocylinder fluid confined in a matrix
formed by a disordered hard sphere. We start this study from the bifurcation analysis of
the integral equation~(\ref{hol2.24}) for the singlet distribution function
$f\left( \Omega\right) $. This equation has the same form as the corresponding equation
obtained by Onsager \cite{2hol} for the hard spherocylinder fluid in the
limit of $L_{1}\to\infty$, $D_{1}\to 0$ while the dimensional density of fluid
 $c_1=\frac{1}{4}\piup \rho_1 L_1^{2}D_1$ is fixed. In the Onsager limit
\begin{align}
C \to c_{1}=\frac{1}{4}\piup \rho_1 L_1^{2}D_1.
\label{hol5.5}
\end{align}

From the bifurcation analysis of the integral equation~(\ref{hol2.24}) it was
found that this equation has two characteristic points $C_{\text i}$ and $C_{\text n}$
\cite{28hol}, which define the range of stability of a considered system. The
first point $C_{\text i}$ corresponds to the highest value of a possible density of a
stable isotropic state and the second point $C_{\text n}$ corresponds to the lowest
value of a possible density of a stable nematic state. For the Onsager model
from the solution of the coexistence equations, the values of the density of
coexisting isotropic and nematic phases were obtained
\cite{29hol,30hol,31hol}
\begin{align}
c_{\text i}=3.289, \qquad c_{\text n}=4.192.
\label{hol5.6}
\end{align}

For the finite values of $L_{1}$ and $D_{1}$ we can put
\begin{align}
C_{\text i}=3.289, \qquad C_{\text n}=4.192.
\label{hol5.7}
\end{align}

For the constant $C$ in this paper we have three different approximations. In the SPT2b1 $C$ is given by the expression~(\ref{hol2.25}), for CS-PL 
approximation $C$ is given by the expression~(\ref{hol3.9}) and for PL approximation $C$ is given by the expression~(\ref{hol4.6}). The values (\ref{hol5.7}) for $C$ define the isotropic-nematic phase diagram
for a hard spherocylinder fluid in disordered porous media depending on
the ratio $L_{1}/D_{1}=\gamma_{1}-1$ and the parameter of the matrix, namely $\eta_{0}=1-\phi_{0}$ and the ratio $D_{1}/D_{0}$. To be more
specific, we will fix the last ratio by putting $D_{0}=L_{1}$. As a result,
$\frac{D_{1}}{D_{0}}=1/(\gamma_{1}-1).$

\begin{figure}[!t]
	\centerline{
	\includegraphics [width=0.58\textwidth]{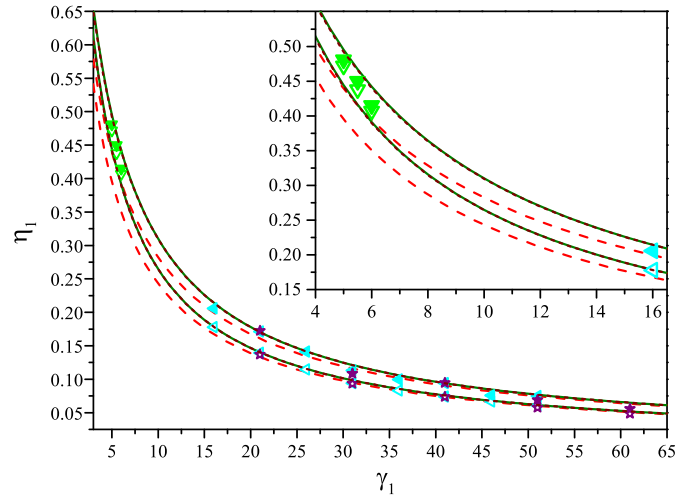}
	}
	\caption{(Colour online) Isotropic-nematic coexistence diagram in the bulk case for a hard spherocylinder fluid in the plane of the packing fraction of the fluid $\eta_{1}$
versus parameter $\gamma_{1}$. The results presented are obtained from the bifurcation analysis of the integral equation~(\ref{hol2.24})
in different approximations for the constant~$C$: red dashed line denotes SPT2b1, green solid line denotes PL, brown dotted line denotes CS-PL,
down pointed green
triangles $\bigtriangledown$ are the results of the simulations \cite{32hol}, left pointed triangles $\lhd$ are GDI simulation results
\cite{33hol} and star points $\star$ are the GEMC simulation results taken from \cite{33hol}.}
	\label{Fig1}
\end{figure}

At the beginning we will demonstrate how the developed approaches describe
the isotropic-nematic coexistence curves for a hard spherocylinder fluid in
the bulk case. In figure~\ref{Fig1} we present the dependence of the density $\eta_1$ on
the parameter $\gamma_{1}$ along the isotropic-nematic coexistence curves
obtained from the bifurcation analysis in SPT2b1, CS-PL and PL
approximations. For comparison the computer simulation results taken from
\cite{32hol,33hol} are presented as well. As we can see all three approximations
for large enough values of $\gamma_{1}$ give the same results in good
agreement with the simulation data. However, starting from $\gamma_{1}$ near
$\gamma_{1}\approx30$ there is a deviation of the SPT2b1 approximation from
general tendency and the computer simulation data. This deviation increases
with a decreasing parameter $\gamma_{1}$ and at $\gamma_{1}$ smaller than 20 it
leads to incorrect results. Two other approximations, namely the CS-PL and PL
ones, reproduce more correctly the general tendency of the dependence of
coexistence curves on $\gamma_{1}$. We do not observe differences between the
CS-PL and PL approximations. However, at small $\gamma_{1}$, the results
predicted from the bifurcation analysis slightly overestimate the jump of
the density at the phase transition.

We should note that the isotropic-nematic coexistence lines can also be found from the condition of thermodynamic equilibrium, according to which the
isotropic and nematic phases have the same pressure and the same chemical potential:
\begin{align}
P\left( \eta_{\text i}\right) =P(\eta_{\text n}), \qquad \mu\left( \eta_{\text i}\right) =\mu_{1}(\eta_{\text n}).
\label{hol5.8}
\end{align}

\begin{figure}[!t]
	\centerline{
		\includegraphics [width=0.58\textwidth]{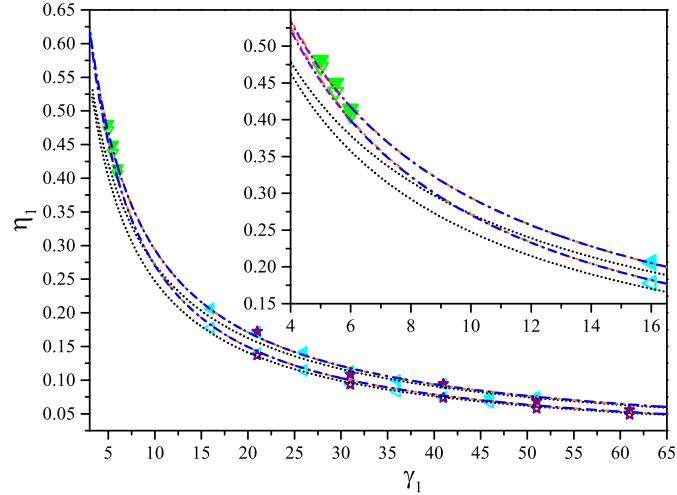}
	}
	\caption{(Colour online) Isotropic-nematic coexistence diagram in the bulk case for a hard spherocylinder fluid in the plane of the packing fraction of the fluid $\eta_{1}$
versus parameter $\gamma_{1}$. The results presented are obtained from the thermodynamics analysis in different approximations: black
dotted line denotes SPT2b1, orange dash-dot-dot line denotes PL, blue dash-dot-dash line denotes CS-PL, down pointed green triangles $\bigtriangledown$ are
the results of simulations taken from \cite{32hol}, left pointed triangles $\lhd$ are GDI simulations results from \cite{33hol} and star points $\star$ are
the GEMC simulations results taken from \cite{33hol}.}
	\label{Fig2}
\end{figure}

The coexistence curves obtained from the condition (\ref{hol5.8}) for a hard
spherocylinder fluid in the bulk case are presented in figure~\ref{Fig2}. As it was
shown in \cite{28hol}, in the Onsager limit, the results obtained from the
bifurcation analysis and from the condition of thermodynamic equilibrium
coincide exactly. Similar to the bifurcation analysis in the
thermodynamic way for large enough values of $\gamma_{1}$ all three
approximations give the same results, but with a decreasing $\gamma_{1}$ we
observed a deviation in SPT2b1 approximation and computer simulation data
which leads to incorrect results at small $\gamma_{1}$. Again we do not
observe the difference between the CS-PL and PL results. However, at
small $\gamma_{1}$, contrary to the bifurcation analysis, the thermodynamic
consideration slightly underestimates the value of the density jump between
the isotropic and nematic phases. Nevertheless, comparing figure~\ref{Fig1} and figure~\ref{Fig2}
we can see that thermodynamic consideration leads to a better agreement with
the computer simulation data.

\begin{figure}[!t]
	\centerline{
		\includegraphics [width=0.58\textwidth]{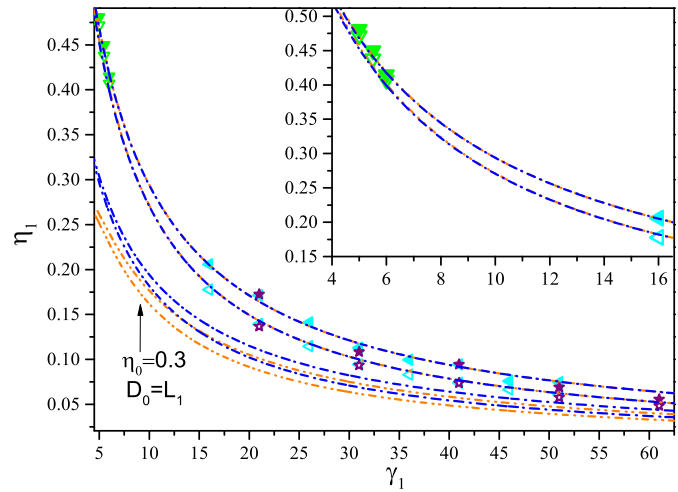}
	}
	\caption{(Colour online) The influence of the porous medium on the isotropic-nematic coexistence diagram for a hard spherocylinder fluid in the plane of the
packing fraction of the fluid $\eta_{1}$ versus parameter $\gamma_{1}$. The results are calculated from the thermodynamics analysis. The results for
a fluid in the disordered porous medium with the porosity $\phi_{0}=0.7$ $(\eta_{0}=0.3)$ are presented by dotted lines. For comparison purposes the
results for the bulk case are also presented. The notations are the same as in figure~\ref{Fig2}.}
	\label{Fig3}
\end{figure}

\begin{figure}[!t]
	\centerline{
		\includegraphics [width=0.57\textwidth]{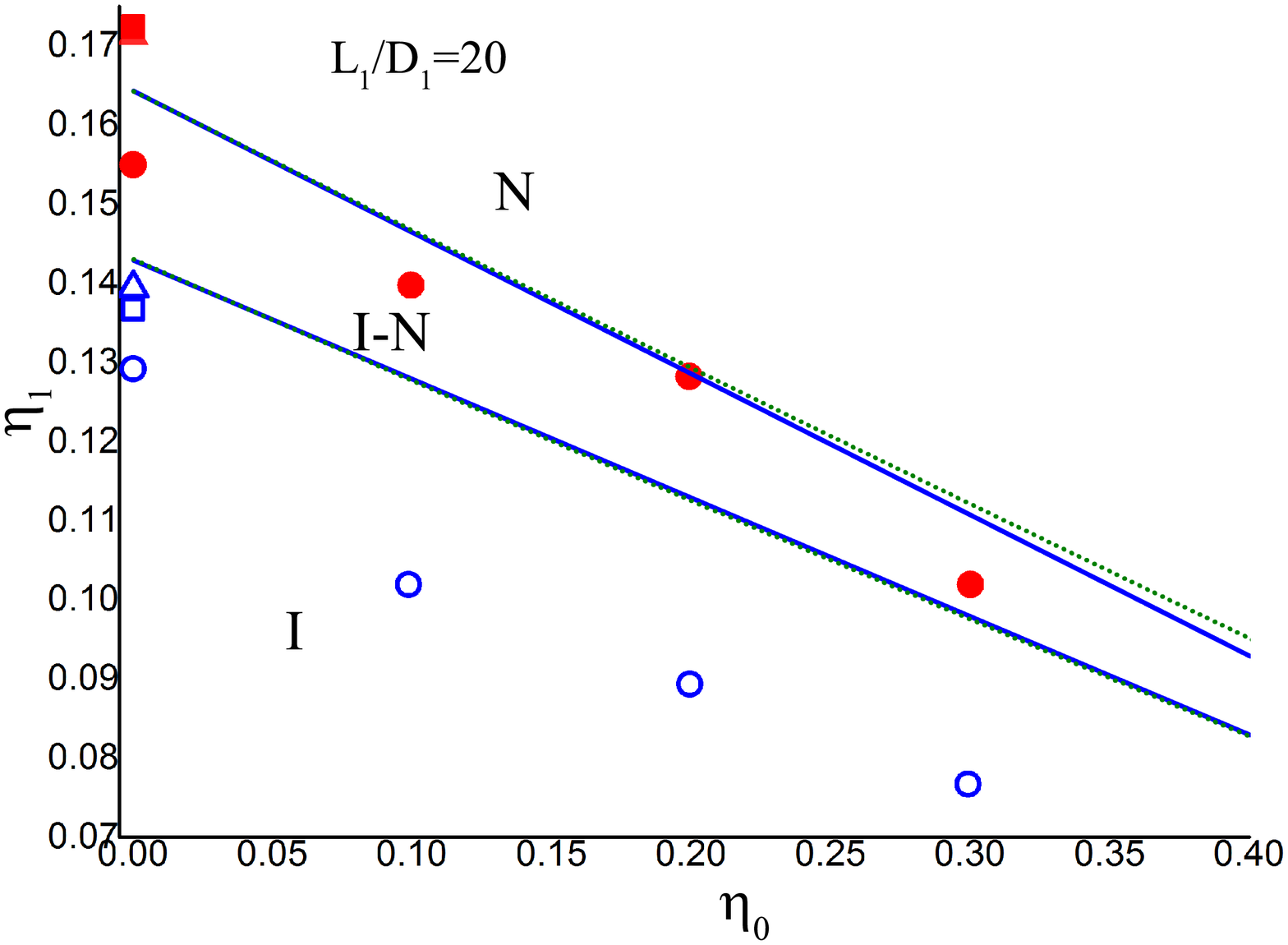}
	}
	\caption{(Colour online) Coexistence lines of isotropic-nematic phases of a hard spherocylinder fluid in a hard sphere matrix for $L_{1}/D_{1}=20$ and
$D_{0}/L_{1}=1$. Dependencies of the spherocylinder fluid packing fraction $\eta_{1}$ on the matrix packing fraction $\eta_{0}$ are presented.
The results are obtained from the thermodynamics analysis with green dotted lines corresponding to the PL approximation and the blue solid lines
corresponding to the CS approximation. The GEMC results taken from
\cite{34hol} are shown as circles and those taken from \cite{33hol} are shown as squares and triangles (GDI).}
	\label{Fig4}
\end{figure}

\begin{figure}[!t]
	\centerline{
		\includegraphics [width=0.57\textwidth]{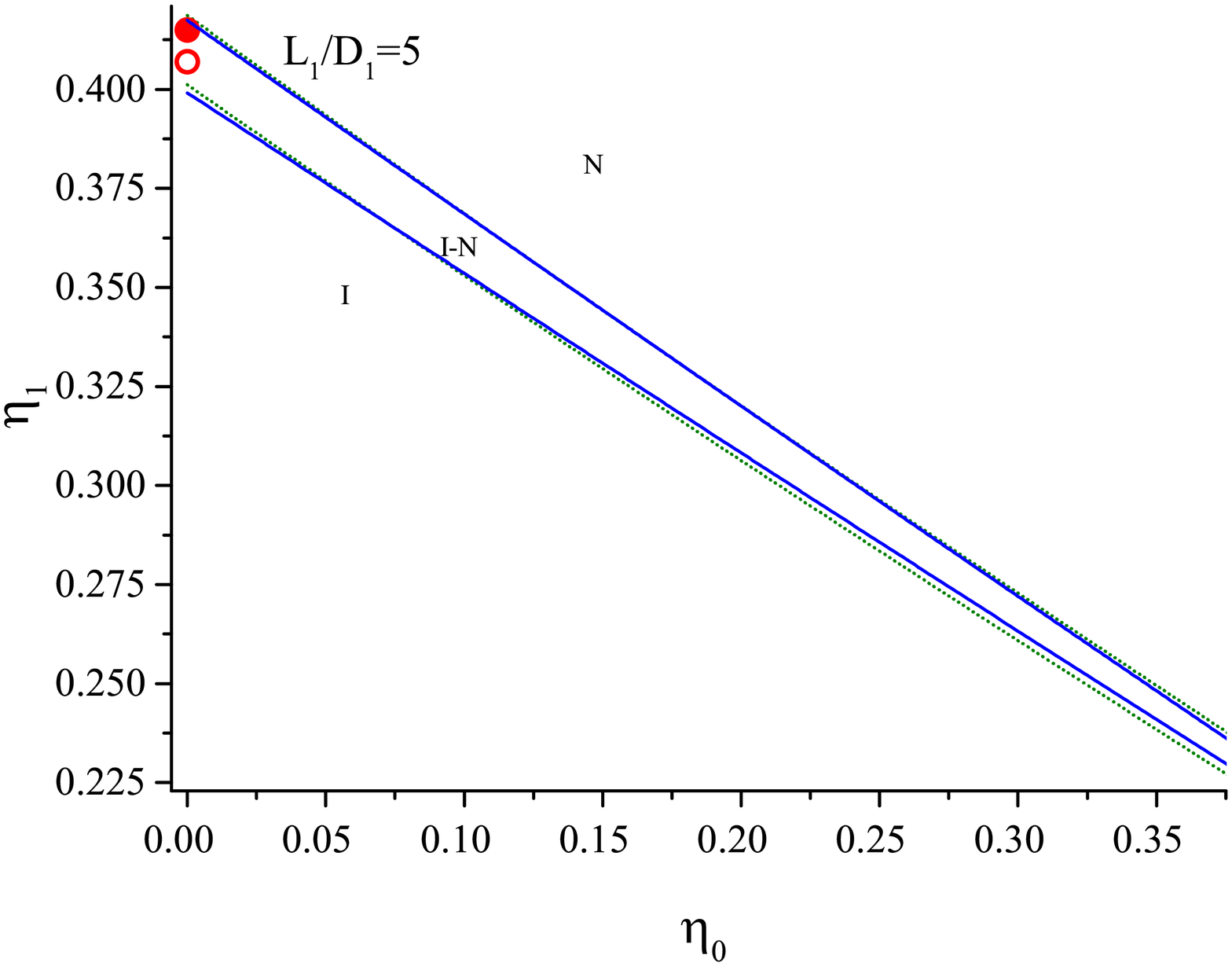}
	}
	\caption{(Colour online) Coexistence lines of isotropic-nematic phases of a hard spherocylinder fluid in a hard sphere matrix for $L_{1}/D_{1}=5$ and
$D_{0}/L_{1}=1$. Dependencies of the spherocylinder fluid packing fraction $\eta_{1}$ on the matrix packing fraction $\eta_{0}$ are presented.
The results are obtained from the thermodynamics analysis with green dotted lines corresponding to the PL approximation and the blue solid lines
corresponding to the CS approximation. The GEMC results taken from
\cite{32hol} are shown as red circles.}
	\label{Fig5}
\end{figure}

As we have already noted in the Onsager limit all three approximations in the
thermodynamic approach and in the bifurcation analysis reproduce correctly
the exact result (\ref{hol5.6}). In the presence of a porous medium for the
Onsager model we obtain
\begin{align}
c_{\text i}/\phi_{0}=3.289, \qquad c_{\text n}/\phi_{0}=4.192.
\label{hol5.9}
\end{align}

It means that for the isotropic-nematic phase transition, the presence of a
porous medium shifts the phase diagram to lower densities of a  fluid. This
effect of the porous medium is illustrated in figure~\ref{Fig3} where the dependence of
the density of fluid  $\eta_{1}$ on parameter $\gamma_{1}$ along the
isotropic-nematic coexistence curves calculated from thermodynamic
equilibrium (\ref{hol5.8}) for the hard spherocylinder fluid in a porous medium with the
porosity $\phi_{0}=0.7$ $(\eta_{0}=0.3)$ is presented. For comparison, the
isotropic-nematic diagram for the bulk case is also presented.

The influence of porosity $\phi_{0}$ on the coexistence lines of the
isotropic-nematic phase transition calculated from the conditions
(\ref{hol5.8}) for a hard spherocylinder fluid in coordinates fluid density $\eta_{1}$ versus
packing fraction $\eta_{0}$ of matrix particles (the porosity
$\phi_{0}=1-\eta_{0}$) is illustrated also in figure~\ref{Fig4} and figure~\ref{Fig5} for the
cases $L_{1}/D_{1}=20$ and $L_{1}/D_{1}=5$, correspondingly. All the curves
are obtained from the condition of thermodynamic equilibrium (\ref{hol5.8}) in
CS-PL and PL approximations, correspondingly. As we can see there are small insignificant
differences between the predictions from both approaches. For comparison, in
figure~\ref{Fig4} the results of the computer simulations of Schmidt and Dijkstra
\cite{34hol} obtained by the method of Gibbs ensemble Monte Carlo (GEMC) are
also presented. For the bulk case $(\eta_{0}=0)$, the
results of Bolhuis and Frenkel \cite{33hol} obtained by GEMC method and GEMC
combined with modified Gibbs-Duhem integration (GDI) method are shown as
well. As we can see there are some differences between the
computer data from \cite{34hol} and \cite{33hol}. Our theoretical prediction is in better agreement with the data from \cite{33hol}  and correctly reproduces the dependences of
$\eta_{1}$ on $\eta_{0}$ along the coexistence curves.  The computer simulation results from \cite{32hol} are also presented in figure~\ref{Fig5}. We see a good correlation between computer simulation data and theoretical prediction.

\section{Conclusions \label{s6}}

In this paper the scaled particle theory (SPT) is extended for the
description of a hard spherocylinder fluid in a disordered porous medium. We
started from the SPT2b1 approach previously developed by us
\cite{21hol,22hol} for a hard sphere fluid in a disordered porous medium and
generalized in \cite{26hol} for a hard spherocylinder fluid. The theory in
this paper is applied for the study of the influence of disordered porous
media on the isotropic-nematic transition in a hard spherocylinder fluid. It
is shown that the accuracy of the SPT2b1 decreases with decreasing lengths of
spherocylinders.  Two different approaches are developed in order to improve the SPT2b1 theory. In one of them, the so-called SPT2b1-CS-PL
approach, two corrections are involved. The first one is the
Carnahan-Starling correction which improves SPT description of thermodynamical properties at higher
densities of the fluid. The second one corrects the description of
orientational ordering in a hard spherocylinder fluid at higher densities.
The constant of this correction is obtained from the comparison of the
integral equation for the singlet distribution function of a hard
spherocylinder fluid in the SPT2b1 and Parsons-Lee (PL) approaches. In the
second approach, the so-called SPT2b1-PL approach, the PL theory \cite{16hol}
is generalized for a hard spherocylinder fluid in a disordered porous medium.
To this end, according to the original PL theory \cite{16hol}, thermodynamic
properties of a hard spherocylinder fluid in a disordered porous medium are
mapped with the thermodynamic properties of a hard sphere fluid in a
disordered porous medium in the SPT2b1 approximation \cite{21hol,22hol} with
the CS correction considered in this paper.

The phase diagram of a hard spherocylinder fluid in a disordered porous
medium is calculated in two different ways. One of them is connected with the
bifurcation analysis of the nonlinear integral equation for the singlet
distribution function obtained from minimization of the free energy of the
considered system. The second way is based on the condition of thermodynamic
equilibrium. The obtained results are compared with the existing computer
simulation data \cite{32hol,33hol,34hol}. It is shown that in both approaches
the original SPT2b1 approximation is not very accurate with the decreasing
length of spherocylinders. The SPT2b1-CS-PL and SPT2b1-PL approximations in
the bifurcation analysis and in thermodynamic way more or less correctly
reproduce the coexistence curves with decreasing lengths of
spherocylinders. We do not find a significant difference between the
SPT2b1-CS-PL and SPT2b1-PL approximations. However, the bifurcation analysis
slightly overestimates the change of density at the phase transition. In
thermodynamic way, we also did not find the best agreement between theoretical
prediction and computer simulation data for small enough lengths of
spherocylinders. In any case, the thermodynamic way is more preferable for
the description of a phase transition. From bifurcation analysis and from
thermodynamic way we show that the porous medium shifts the phase diagram to
lower densities of a fluid. Comparison with computer simulation results is
discussed.

The model considered in this paper can be used as the reference system for
generalization of the Van der Waals theory for anisotropic fluids in
disordered porous media \cite{26hol,27hol,35hol} and for taking 
different types of associations \cite{36hol} into account.

\ukrainianpart

\title{Теорія масштабної частинки для системи сфероциліндричного плину в невпорядкованому пористому середовищі: поправки Карнагана-Старлінга і Парсонса-Лі}
\author{М.Ф. Головко, В.І. Шмотолоха}
\address{
 Iнститут фiзики конденсованих систем НАН України, вул. I. Свєнцiцького, 1, 79011 Львiв, Україна
}

\makeukrtitle

\begin{abstract}

Теорія масштабної частинки (ТМЧ) застосовується для вивчення впливу пористого середовища на ізотропно-нематичний перехід у плині твердих сфероциліндрів. Розроблено два нові підходи для покращення опису сфероциліндрів невеликої довжини. В одному з них, так званому підході ТМЧ-КС-ПЛ, вводиться поправка Карнагана-Старлінга (КС) для покращення опису термодинамічних властивостей плину, тоді як поправка Парсонса-Лі (ПЛ) покращує
опис орієнтаційного впорядкування. Другий підхід, так званий підхід ТМЧ-ПЛ, пов'язаний з узагальненням теорії Парсонса-Лі для анізотропних рідин у 	невпорядкованих пористих середовищах. Фазова діаграма отримана з біфуркаційного аналізу нелінійного інтегрального рівняння для одночастинкової функції розподілу та умови термодинамічної рівноваги. Отримані дані порівнюються з даними комп'ютерних симуляцій. Обидва шляхи і обидва підходи істотно покращують опис системи сфероциліндричного плину у випадку малих довжин сфероциліндра. Ми не знайшли істотної різниці в результатах в обох розроблених підходах. Ми виявили, що біфуркаційний аналіз трохи переоцінює, а термодинамічний аналіз недооцінює передбачення, отримані з комп'ютерних симуляцій. Пористе середовище зсуває фазову діаграму в бік менших густин плину і не змінює тип переходу.
	\tolerance=3000%

\keywords твердий сфероциліндричний плин, пористий матеріал, теорія масштабної частинки,
	ізотропно-нематичний перехід, теорія Парсонса-Лі, поправка Карнагана-Старлінга
	
\end{abstract}

\end{document}